\input epsf
\input amssym 
\catcode`@=11                                   
\catcode`\|=12                                  
\catcode`\&=4                                   

\newcount\ncols         \ncols=\z@              
\newcount\nrows         \nrows=\z@              
\newcount\curcol        \curcol=\z@             
     
\newdimen\thinsize      \thinsize=0.6pt         
\newdimen\thicksize     \thicksize=1.5pt        

\newif\iftableinfo      \tableinfotrue          
\newif\ifcentertables   \centertablestrue       
%
%
     
\let\plaincr=\cr                        
\let\plainspan=\span                    
\let\plaintab=&                         
\let\lparen=(                           
\let\NX=\noexpand                       

     
\def\ruledtable{\relax                          
    \@BeginRuledTable                           
    \@RuledTable}


\def\@BeginRuledTable{
   \ncols=0\nrows=0                             
   \begingroup                                  
    \offinterlineskip                           
    \def~{\phantom{0}}
    \def\span{\plainspan\omit\relax\colcount\plainspan}
    \let\cr=\crrule                             
    \let\CR=\crthick                            
    \let\nr=\crnorule                           
    \let\|=\Vb                                  
%
%
    \ifx\tablestrut\undefined\relax             
    \else\let\tstrut=\tablestrut\fi             
    \catcode`\|=13 \catcode`\&=13\relax         
    \TableActive                                
    \curcol=1                                   
%
%
    \ifdim\tablewidth>-\maxdimen\relax          %
      \edef\@Halign{\NX\halign to \NX\tablewidth\NX\bgroup\TablePreamble}%
      \tabskip=0pt plus 1fil                    
    \else                                       %
      \edef\@Halign{\NX\halign\NX\bgroup\TablePreamble}%
      \tabskip=0pt                              
    \fi                                         %
%
%
    \ifcentertables                             
       \ifhmode\vskip 0pt\fi                    
       \line\bgroup\hss                         
    \else\hbox\bgroup                           
    \fi}


\long\def\@RuledTable#1\endruledtable{
   \vrule width\thicksize                       
     \vbox{\@Halign                             
       \thickrule                               
       #1\relax                                 
       \tstrut                                  
       \plaincr\thickrule                       
     \egroup}
   \vrule width\thicksize                       
   \ifcentertables\hss\fi\egroup                
  \endgroup                                     
  \global\tablewidth=-\maxdimen                 
  \iftableinfo                                  
      \immediate\write16{[Nrows=\the\nrows, Ncols=\the\ncols]}%
   \fi}
     

\def\TablePreamble{
   \linecount                           
   \TableItem{####}
   \plaintab\plaintab                   
   \TableItem{####}
   \plaincr}


\def\@TableItem#1{
   \hfil\tablespace                             
   #1\relax                                     
   \tablespace\hfil                             
    }%

\def\@tableright#1{
   \hfil\tablespace\relax               
   #1\relax                             
   \tablespace\relax}

\def\@tableleft#1{
   \tablespace\relax                    
   #1\relax                             
   \tablespace\hfil}

\let\TableItem=\@TableItem              
     
\def\RightJustifyTables{\let\TableItem=\@tableright}
\def\LeftJustifyTables{\let\TableItem=\@tableleft}
\def\NoJustifyTables{\let\TableItem=\@TableItem}

\def\LooseTables{\let\tablespace=\quad}
\def\TightTables{\let\tablespace=\space}
\LooseTables                                    

%

\newdimen\tablewidth    \tablewidth=-\maxdimen  


\def\setRuledStrut{
   \dimen@=\baselineskip                        
   \advance\dimen@ by-\normalbaselineskip       
   \ifdim\dimen@<.5ex \dimen@=.5ex\fi           
   \setbox0=\hbox{\lparen}
   \dimen1=\dimen@ \advance\dimen1 by \ht0      
   \dimen2=\dimen@ \advance\dimen2 by \dp0      
   \def\tstrut{\vrule height\dimen1 depth\dimen2 width\z@}%
   }%

\def\tstrut{\vrule height 3.1ex depth 1.2ex width 0pt}


\def\bigitem#1{
   \setbox0=\hbox{#1}
   \dimen1 =\ht0 \dimen2 =\dp0                  
   \dimen@ =\baselines@ve                       
   \advance\dimen@ by-\normalbaselineskip       
   \ifdim\dimen@<.25ex \dimen@=.25ex\fi         
   \advance\dimen1 by \dimen@                   
   \advance\dimen2 by \dimen@                   
   \vrule height\dimen1 depth\dimen2 width\z@   
   \copy0}

     
%

     
\def\nextcolumn#1{
   \plaintab\omit#1\relax\colcount              
   \plaintab}
     
\def\tab{
   \nextcolumn{\relax}}


\def\vb{
   \nextcolumn{\vrule width\thinsize}}

\def\Vb{
   \nextcolumn{\vrule width\thicksize}}

\def\dbl{
   \nextcolumn{\vrule width\thinsize            
   \hskip\thinsize\vrule width\thinsize}}

     
{\catcode`\|=13 \let|0
 \catcode`\&=13 \let&0
 \gdef\TableActive{\let|=\vb \let&=\tab}%
}


\def\crrule{\relax                      
   \tstrut                              
   \plaincr\tablerule                   
  }%

\def\crthick{\relax                     
   \tstrut                              
   \plaincr\thickrule                   
  }%
     
\def\crnorule{\relax                    
   \tstrut                              
   \plaincr                             
   }%
   

     
\def\tablerule{\noalign{\hrule height\thinsize depth 0pt}}%
\def\thickrule{\noalign{\hrule height\thicksize depth 0pt}}%


%
%
%
     

\def\linecount{\relax\global\ncols=\curcol      
   \global\curcol=1                             
   \global\advance\nrows by 1\relax}
     
\def\colcount{\relax                            %
   \global\advance\curcol by 1\relax}


\newdimen\parasize      \parasize=4in           

%

%

\def\begintable{\relax                          
    \@BeginRuledTable                           
    \@begintable}

\long\def\@begintable#1\endtable{
   \@RuledTable#1\endruledtable}


\catcode`@=12                                   



\newfam\scrfam
\batchmode\font\tenscr=rsfs10 \errorstopmode
\ifx\tenscr\nullfont
        \message{rsfs script font not available. Replacing with calligraphic.}
        \def\scr{\cal}
\else   
        \font\sevenscr=rsfs7
        \font\fivescr=rsfs5
        \skewchar\tenscr='177 \skewchar\sevenscr='177 \skewchar\fivescr='177
        \textfont\scrfam=\tenscr \scriptfont\scrfam=\sevenscr
        \scriptscriptfont\scrfam=\fivescr
        \def\scr{\fam\scrfam}
        \def\cal{\scr}
\fi
\catcode`\@=11
\newfam\frakfam
\batchmode\font\tenfrak=eufm10 \errorstopmode
\ifx\tenfrak\nullfont
        \message{eufm font not available. Replacing with italic.}
        
\else
    
    \font\sevenfrak=eufm7 \font\fivefrak=eufm5
    \textfont\frakfam=\tenfrak
    \scriptfont\frakfam=\sevenfrak \scriptscriptfont\frakfam=\fivefrak
    
\fi
\catcode`\@=\active
\newfam\msbfam
\batchmode\font\twelvemsb=msbm10 scaled\magstep1 \errorstopmode
\ifx\twelvemsb\nullfont\def\Bbb{\bf}

    \message{Blackboard bold not available. Replacing with boldface.}
\else   \catcode`\@=11
        \font\tenmsb=msbm10 \font\sevenmsb=msbm7 \font\fivemsb=msbm5
        \textfont\msbfam=\tenmsb
        \scriptfont\msbfam=\sevenmsb \scriptscriptfont\msbfam=\fivemsb
        \def\Bbb{\relax\expandafter\Bbb@}
        \def\Bbb@#1{{\Bbb@@{#1}}}
        \def\Bbb@@#1{\fam\msbfam\relax#1}
        \catcode`\@=\active

\fi
\newfam\cpfam
\def\sectionfonts{\relax
    \textfont0=\twelvecp          \scriptfont0=\ninecp
      \scriptscriptfont0=\sixrm
    \textfont1=\twelvei           \scriptfont1=\ninei
      \scriptscriptfont1=\sixi
    \textfont2=\twelvesy           \scriptfont2=\ninesy
      \scriptscriptfont2=\sixsy
    \textfont3=\twelveex          \scriptfont3=\tenex
      \scriptscriptfont3=\tenex
    \textfont\itfam=\twelveit     \scriptfont\itfam=\nineit
    \textfont\slfam=\twelvesl     \scriptfont\slfam=\ninesl
    \textfont\bffam=\twelvebf     \scriptfont\bffam=\ninebf
      \scriptscriptfont\bffam=\sixbf
    \textfont\ttfam=\twelvett
    \textfont\cpfam=\twelvecp
}
        \font\eightrm=cmr8              \def\xrm{\eightrm}
        \font\eightbf=cmbx8             \def\xbf{\eightbf}
        \font\eightit=cmti10 at 8pt     \def\xit{\eightit}
                       
        \font\sixrm=cmr6                
        \font\eighttt=cmtt8             
        \font\eightcp=cmcsc8
        \font\eighti=cmmi8              \def\xold{\eighti}
        \font\eightib=cmmib8             \def\xbold{\eightib}
        \font\teni=cmmi10               \def\old{\teni}
        \font\ninei=cmmi9
        \font\tencp=cmcsc10
        \font\ninecp=cmcsc9

        \font\twelvei=cmmi12
        \font\twelvecp=cmcsc10 scaled\magstep1
        
        \font\fiverm=cmr5
        
        \font\twelvesy=cmsy12
        \font\ninesy=cmsy9
        \font\sixsy=cmsy6
        \font\twelveex=cmex12

        \font\twelveit=cmti12
        \font\nineit=cmti9
        
        \font\twelvesl=cmsl12
        \font\ninesl=cmsl9
        
        \font\twelvebf=cmbx12
        \font\ninebf=cmbx9
        \font\sixbf=cmbx6
        \font\twelvett=cmtt12

        \font\sixi=cmmi6

\batchmode\font\tenhelvbold=phvb at10pt \errorstopmode
\ifx\tenhelvbold\nullfont
        \message{phvb font not available. Replacing with cmr.}
    \font\tenhelvbold=cmb10   
    \font\twelvehelvbold=cmb12
    
    \font\sixteenhelvbold=cmb16
  \else
    \font\tenhelvbold=phvb at10pt   
    \font\twelvehelvbold=phvb at12pt
     at14pt
    \font\sixteenhelvbold=phvb at16pt
\fi

\def\noblackbox{\overfullrule=0pt}
\noblackbox

\font\eightmi=cmmi8
\font\sixmi=cmmi6
\font\fivemi=cmmi5

\font\eightsy=cmsy8
\font\sixsy=cmsy6
\font\fivesy=cmsy5

\font\eightsl=cmsl8

\def\eightpoint{
\def\rm{\fam0\eightrm}
\textfont0=\eightrm \scriptfont0=\sixrm \scriptscriptfont0=\fiverm
\textfont1=\eightmi  \scriptfont1=\sixmi  \scriptscriptfont1=\fivemi
\textfont2=\eightsy \scriptfont2=\sixsy \scriptscriptfont2=\fivesy
\textfont3=\tenex   \scriptfont3=\tenex \scriptscriptfont3=\tenex
\textfont\itfam=\eightit \def\it{\fam\itfam\eightit}
\textfont\slfam=\eightsl \def\sl{\fam\slfam\eightsl}
\textfont\ttfam=\eighttt \def\tt{\fam\ttfam\eighttt}
\textfont\bffam=\eightbf \scriptfont\bffam=\sixbf 
                         \scriptscriptfont\bffam=\fivebf
                         \def\bf{\fam\bffam\eightbf}
\normalbaselineskip=10pt}

\newtoks\headtext
\headline={\ifnum\pageno=1\hfill\else
    \ifodd\pageno{\eightcp\the\headtext}{ }\dotfill{ }{\old\folio}
    \else{\old\folio}{ }\dotfill{ }{\eightcp\the\headtext}\fi
    \fi}
\def\makeheadline{\vbox to 0pt{\vss\noindent\the\headline\break
\hbox to\hsize{\hfill}}
        \vskip2\baselineskip}
\newcount\infootnote
\infootnote=0
\def\foot#1#2{\infootnote=1
\footnote{${}^{#1}$}{\vtop{\baselineskip=.75\baselineskip
\advance\hsize by
-\parindent{\eightpoint\rm\hskip-\parindent #2}\hfill\vskip\parskip}}\infootnote=0$\,$}
\newcount\refcount
\refcount=1
\newwrite\refwrite
\def\oldsize{\ifnum\infootnote=1\xold\else\old\fi}
\def\ref#1#2{
    \def#1{{{\oldsize\the\refcount}}\ifnum\the\refcount=1\immediate\openout\refwrite=\jobname.refs\fi\immediate\write\refwrite{\item{[{\xold\the\refcount}]}
    #2\hfill\par\vskip-2pt}\xdef#1{{\noexpand\oldsize\the\refcount}}\global\advance\refcount by 1}
    }
\def\refout{\catcode`\@=11
        \xrm\immediate\closeout\refwrite
        \vskip2\baselineskip
        {\noindent\twelvecp References}\hfill
        \par\nobreak\vskip\baselineskip
        \baselineskip=.75\baselineskip
        \input\jobname.refs
        \baselineskip=4\baselineskip \divide\baselineskip by 3
        \catcode`\@=\active\rm}

\def\hepth#1{\href{http://arxiv.org/abs/hep-th/#1}{arXiv:hep-th/{\xold#1}}}

\def\arxiv#1#2{\href{http://arxiv.org/abs/#1.#2}{arXiv:{\xold#1}.{\xold#2}}}
\def\jhep#1#2#3#4{\href{http://jhep.sissa.it/stdsearch?paper=#2\%28#3\%29#4}{J. High Energy Phys. {\xbold #1#2} ({\xold#3}) {\xold#4}}}

\def\CQG#1#2#3{Class. Quantum Grav. {\xbold#1} ({\xold#2}) {\xold#3}}

\def\JHEP{\jhep}

\def\NPB#1#2#3{Nucl. Phys. {\xbf B}{\xbold#1} ({\xold#2}) {\xold#3}}

\def\PLB#1#2#3{Phys. Lett. {\xbf B}{\xbold#1} ({\xold#2}) {\xold#3}}

\def\PRD#1#2#3{Phys. Rev. {\xbf D}{\xbold#1} ({\xold#2}) {\xold#3}}

\newcount\sectioncount
\sectioncount=0
\def\section#1#2{\global\eqcount=0
    \global\subsectioncount=0
        \global\advance\sectioncount by 1
    \ifnum\sectioncount>1
            \vskip2\baselineskip
    \fi
    \noindent
       \line{\sectionfonts\twelvecp\the\sectioncount. #2\hfill}
        \par\nobreak\vskip.8\baselineskip\noindent
        \xdef#1{{\old\the\sectioncount}}}
\newcount\subsectioncount
\def\subsection#1#2{\global\advance\subsectioncount by 1
    \par\nobreak\vskip.8\baselineskip\noindent
    \line{\tencp\the\sectioncount.\the\subsectioncount. #2\hfill}
    \vskip.5\baselineskip\noindent
    \xdef#1{{\old\the\sectioncount}.{\old\the\subsectioncount}}}
\newcount\appendixcount
\appendixcount=0
\def\appendix#1{\global\eqcount=0
        \global\advance\appendixcount by 1
        \vskip2\baselineskip\noindent
        \ifnum\the\appendixcount=1
        \hbox{\twelvecp Appendix A: #1\hfill}
        \par\nobreak\vskip\baselineskip\noindent\fi
    \ifnum\the\appendixcount=2
        \hbox{\twelvecp Appendix B: #1\hfill}
        \par\nobreak\vskip\baselineskip\noindent\fi
    \ifnum\the\appendixcount=3
        \hbox{\twelvecp Appendix C: #1\hfill}
        \par\nobreak\vskip\baselineskip\noindent\fi}
\def\acknowledgements{\vskip2\baselineskip\noindent
        \underbar{\it Acknowledgements:}\ }
\newcount\eqcount
\eqcount=0
\def\Eqn#1{\global\advance\eqcount by 1
\ifnum\the\sectioncount=0
    \xdef#1{{\old\the\eqcount}}
    \eqno({\oldstyle\the\eqcount})
\else
        \ifnum\the\appendixcount=0
            \xdef#1{{\old\the\sectioncount}.{\old\the\eqcount}}
                \eqno({\oldstyle\the\sectioncount}.{\oldstyle\the\eqcount})\fi
        \ifnum\the\appendixcount=1
            \xdef#1{{\oldstyle A}.{\old\the\eqcount}}
                \eqno({\oldstyle A}.{\oldstyle\the\eqcount})\fi
        \ifnum\the\appendixcount=2
            \xdef#1{{\oldstyle B}.{\old\the\eqcount}}
                \eqno({\oldstyle B}.{\oldstyle\the\eqcount})\fi
        \ifnum\the\appendixcount=3
            \xdef#1{{\oldstyle C}.{\old\the\eqcount}}
                \eqno({\oldstyle C}.{\oldstyle\the\eqcount})\fi
\fi}
\def\eqn{\global\advance\eqcount by 1
\ifnum\the\sectioncount=0
    \eqno({\oldstyle\the\eqcount})
\else
        \ifnum\the\appendixcount=0
                \eqno({\oldstyle\the\sectioncount}.{\oldstyle\the\eqcount})\fi
        \ifnum\the\appendixcount=1
                \eqno({\oldstyle A}.{\oldstyle\the\eqcount})\fi
        \ifnum\the\appendixcount=2
                \eqno({\oldstyle B}.{\oldstyle\the\eqcount})\fi
        \ifnum\the\appendixcount=3
                \eqno({\oldstyle C}.{\oldstyle\the\eqcount})\fi
\fi}
\def\multi{\global\advance\eqcount by 1}
\def\multieq#1#2{
    \ifnum\the\sectioncount=0
        \eqno({\oldstyle\the\eqcount})
         \xdef#1{{\old\the\eqcount#2}}
    \else
        \xdef#1{{\old\the\sectioncount}.{\old\the\eqcount}#2}
        \eqno{({\oldstyle\the\sectioncount}.{\oldstyle\the\eqcount}#2)}
    \fi}

\newtoks\url
\def\Href#1#2{\catcode`\#=12\url={#1}\catcode`\#=\active#2}
\def\href#1#2{{#2}}
\def\hhref#1{{#1}}
\parskip=3.5pt plus .3pt minus .3pt
\baselineskip=14pt plus .1pt minus .05pt
\lineskip=.5pt plus .05pt minus .05pt
\lineskiplimit=.5pt
\abovedisplayskip=18pt plus 4pt minus 2pt
\belowdisplayskip=\abovedisplayskip
\hsize=14cm
\vsize=20.5cm
\hoffset=1.5cm
\voffset=1.5cm
\frenchspacing
\footline={}
\raggedbottom

\def\ss{\scriptstyle}
\def\sss{\scriptscriptstyle}
\def\*{\partial}
\def\punkt{\,\,.}
\def\komma{\,\,,}

\def\={\!=\!}
\def\small#1{{\hbox{$#1$}}}

\def\fraction#1{\small{1\over#1}}
\def\fr{\fraction}
\def\Fraction#1#2{\small{#1\over#2}}
\def\Fr{\Fraction}

\def\eg{{\tenit e.g.}}

\def\ie{{\tenit i.e.}}

\def\a{\alpha}
\def\b{\beta}

\def\d{\delta}
\def\e{\varepsilon}
\def\g{\gamma}

\def\m{\mu}





\def\l{\lambda}

\def\RR{{\Bbb R}}

\def\lra{\longrightarrow}

\def\arrowunder#1{\raise4pt\vtop{\baselineskip=0pt\lineskip=0pt
      \ialign{\hfill##\hfill\cr${\ss #1}$\cr$\lra$\cr}}}

\def\xadj{\hbox{\sixbf adj}}

\def\xR{\hbox{\sixbf R}}

\def\leftbr{[\hskip-1.5pt[}
\def\rightbr{]\hskip-1.5pt]}

\def\<{{<}}
\def\>{{>}}

\def\TableItem#1{
   \hfil\tablespace                             
   #1\relax                                     
   \tablespace                             
    }%
\thicksize=\thinsize


\ref\BaggerLambertI{J. Bagger and N. Lambert, {\xit ``Modeling
multiple M2's''}, \PRD{75}{2007}{045020} [\hepth{0611108}].}

\ref\BaggerLambertII{J. Bagger and N. Lambert, {\xit ``Gauge symmetry
and supersymmetry of multiple M2-branes''}, \PRD{77}{2008}{065008}
[\arxiv{0711}{0955}].} 

\ref\BaggerLambertIII{J. Bagger and N. Lambert, {\xit ``Comments on
multiple M2-branes''}, \JHEP{08}{02}{2008}{105} [\arxiv{0712}{3738}].}

\ref\Gustavsson{A. Gustavsson, {\xit ``Algebraic structures on
parallel M2-branes''}, \arxiv{0709}{1260}.}

\ref\Papadopoulos{G. Papadopoulos, {\xit ``M2-branes, 3-Lie algebras
and Pl\"ucker relations''}, \jhep{08}{05}{2008}{054}
[\arxiv{0804}{2662}].}

\ref\GauntlettGutowski{J.P. Gauntlett and J.B. Gutowski, {\xit
``Constraining maximally supersymmetric membrane actions''},
\hfill\break\arxiv{0804}{3078}.} 

\ref\LambertTong{N. Lambert and D. Tong, {\xit ``Membranes on an
orbifold''}, \arxiv{0804}{1114}.}

\ref\DMPvR{J. Distler, S. Mukhi, C. Papageorgakis and M. van
Raamsdonk, {\xit ``M2-branes on M-folds''}, \jhep{08}{05}{2008}{038}
[\arxiv{0804}{1256}].}

\ref\StringTalks{Talks by N. Lambert, J. Maldacena and S. Mukhi at
Strings 2008, CERN, Gen\`eve, August 2008, \hhref{www.cern.ch/strings2008}.}

\ref\CederwallNilssonTsimpisI{M. Cederwall, B.E.W. Nilsson and D. Tsimpis,
{\xit ``The structure of maximally supersymmetric super-Yang--Mills
theory---constraining higher order corrections''},
\jhep{01}{06}{2001}{034} 
[\hepth{0102009}].}

\ref\CederwallNilssonTsimpisII{M. Cederwall, B.E.W. Nilsson and D. Tsimpis,
{\xit ``D=10 super-Yang--Mills at $\ss O(\a'^2)$''},
\JHEP{01}{07}{2001}{042} [\hepth{0104236}].}

\ref\BerkovitsParticle{N. Berkovits, {\xit ``Covariant quantization of
the superparticle using pure spinors''}, \jhep{01}{09}{2001}{016}
[\hepth{0105050}].}

\ref\SpinorialCohomology{M. Cederwall, B.E.W. Nilsson and D. Tsimpis,
{\xit ``Spinorial cohomology and maximally supersymmetric theories''},
\jhep{02}{02}{2002}{009} [\hepth{0110069}];
M. Cederwall, {\xit ``Superspace methods in string theory, supergravity and gauge theory''}, Lectures at the XXXVII Winter School in Theoretical Physics ``New Developments in Fundamental Interactions Theories'',  Karpacz, Poland,  Feb. 6-15, 2001, \hepth{0105176}.}

\ref\Movshev{M. Movshev and A. Schwarz, {\xit ``On maximally
supersymmetric Yang--Mills theories''}, \NPB{681}{2004}{324}
[\hepth{0311132}].}

\ref\BerkovitsI{N. Berkovits,
{\xit ``Super-Poincar\'e covariant quantization of the superstring''},
\jhep{00}{04}{2000}{018} [\hepth{0001035}].}

\ref\BerkovitsNonMinimal{N. Berkovits,
{\xit ``Pure spinor formalism as an N=2 topological string''},
\jhep{05}{10}{2005}{089} [\hepth{0509120}].}

\ref\CederwallNilssonSix{M. Cederwall and B.E.W. Nilsson, {\xit ``Pure
spinors and D=6 super-Yang--Mills''}, \arxiv{0801}{1428}.}

\ref\CGNN{M. Cederwall, U. Gran, M. Nielsen and B.E.W. Nilsson,
{\xit ``Manifestly supersymmetric M-theory''},
\JHEP{00}{10}{2000}{041} [\hepth{0007035}];
{\xit ``Generalised 11-dimensional supergravity''}, \hepth{0010042}.
}

\ref\CGNT{M. Cederwall, U. Gran, B.E.W. Nilsson and D. Tsimpis,
{\xit ``Supersymmetric corrections to eleven-dimen\-sional supergravity''},
\jhep{05}{05}{2005}{052} [\hepth{0409107}].}

\ref\HoweTsimpis{P.S. Howe and D. Tsimpis, {\xit ``On higher order
corrections in M theory''}, \jhep{03}{09}{2003}{038} [\hepth{0305129}].}

\ref\NilssonPure{B.E.W.~Nilsson,
{\xit ``Pure spinors as auxiliary fields in the ten-dimensional
supersymmetric Yang--Mills theory''},
\CQG3{1986}{{\xrm L}41}.}

\ref\HowePureI{P.S. Howe, {\xit ``Pure spinor lines in superspace and
ten-dimensional supersymmetric theories''}, \PLB{258}{1991}{141}.}

\ref\HowePureII{P.S. Howe, {\xit ``Pure spinors, function superspaces
and supergravity theories in ten and eleven dimensions''},
\PLB{273}{1991}{90}.} 

\ref\FreGrassi{P. Fr\'e and P.A. Grassi, {\xit ``Pure spinor formalism
for OSp(N$\ss |$4) backgrounds''}, \arxiv{0807}{0044}.}

\ref\CederwallBLG{M. Cederwall, {\xit ``N=8 superfield formulation of
the Bagger--Lambert--Gustavsson model''}, \arxiv{0808}{3242}.}

\ref\BandosBLG{I. Bandos, {\xit ``NB BLG model in N=8 superfields''},
\arxiv{0808}{3562}.} 

\ref\GranNilssonPetersson{U. Gran, B.E.W. Nilsson and C. Petersson,
{\xit ``On relating multiple M2 and D2-branes''}, \arxiv{0804}{1784}.}

\ref\MarneliusOgren{R. Marnelius and M. \"Ogren, {\xit ``Symmetric
inner products for physical states in BRST quantization''},
\NPB{351}{1991}{474}.} 

\ref\NilssonPalmkvist{B.E.W. Nilsson and J. Palmkvist, {\xit
``Superconformal M2-branes and generalized Jordan triple systems''},
\arxiv{0807}{5134}.} 

\ref\BaggerLambertIV{J. Bagger and N. Lambert, {\xit ``Three-algebras
and N=6 Chern--Simons gauge theories''}, \arxiv{0807}{0163}.}

\ref\BedfordBerman{J. Bedford and D. Berman, {\xit ``A note on quantum
aspects of multiple membranes''}, \arxiv{0806}{4900}.}

\ref\GustavssonII{A. Gustavsson, {\xit ``One-loop corrections to
Bagger--Lambert theory''}, \arxiv{0805}{4443}.}

\ref\BerkovitsICTP{N. Berkovits, {\xit ``ICTP lectures on covariant
quantization of the superstring''}, proceedings of the ICTP Spring
School on Superstrings and Related Matters, Trieste, Italy, 2002
[\hepth{0209059}.]} 

\ref\BandosTownsend{I. Bandos and P.K. Townsend, {\xit ``Light-cone M5
and multiple M2-branes''}, \arxiv{0806}{4777}; {\xit ``SDiff gauge
theory and the M2 condensate''}, \arxiv{0808}{1583}.}

\ref\HoMatsuo{P.-M. Ho and Y. Matsuo, {\xit ``M5 from M2''},
\jhep{08}{06}{2008}{105} [\arxiv{0804}{3629}].}

\ref\SchnablTachikawa{M. Schnabl and Y. Tachikawa, {\xit ``Classification of
superconformal theories of ABJM type''}, \arxiv{0807}{1102}.}

\ref\ABJM{O. Aharony, O. Bergman, D.L. Jafferis and J. Maldacena,
{\xit ``N=6 superconformal Chern--Simons-matter theories, M2-branes
and their gravity duals''}, \arxiv{0806}{1218}.}

\ref\GomisMilanesiRusso{J. Gomis, G. Milanesi and J.G. Russo, {\xit
``Bagger--Lambert theory for general Lie algebras''}, \arxiv{0805}{1012}.}

\ref\BRTV{S. Benvenuti, D. Rodriguez-Gomez, E. Tonni and H. Verlinde,
{\xit ``N=8 superconformal gauge theories and M2 branes''}, 
\arxiv{0805}{1087}.}

\ref\HoImamuraMatsuo{P.-M. Ho, Y. Imamura and Y. Matsuo, {\xit ``M2 to
D2 revisited''}, \jhep{08}{07}{2008}{003} [\arxiv{0805}{1202}].}

\ref\BKKS{M. Benna, I. Klebanov, T. Klose and M. Smedb\"ack, {\xit
``Superconformal Chern--Simons theories and 
AdS${}_{\sss 4}$/ CFT${}_{\sss 3}$
correspondence''}, \arxiv{0806}{1519}.}

\ref\NishiokaTakayanagi{T. Nishioka and T. Takayanagi, {\xit ``On type
IIA Penrose limit and N=6 Chern--Simons theories''}, 
\jhep{08}{08}{2008}{001} [\arxiv{0806}{3391}].}

\ref\MinahanZarembo{J. Minahan and K. Zarembo, {\xit ``The Bethe
Ansatz for superconformal Chern-Simons''}, \arxiv{0806}{3951}.}

\ref\GaiottoWitten{D. Gaiotto and E. Witten, {\xit ``Janus
configurations, Chern--Simons couplings, and the theta-angle in N=4
super-Yang--Mills theory''}, \arxiv{0804}{2907}.}

\ref\HLLLPI{K. Hosomichi, K.-M. Lee, S. Lee, S. Lee and J. Park, {\xit
``N=4 superconformal Chern--Simons theories with hyper and twisted
hyper multiplets''}, \arxiv{0805}{3662}.}

\ref\HLLLPII{K. Hosomichi, K.-M. Lee, S. Lee, S. Lee and J. Park, {\xit
``N=5,6 superconformal Chern--Simons theories and M2-branes on
orbifolds''}, \arxiv{0806}{4977}.} 

\ref\MauriPetkou{A. Mauri and A.C. Petkou, {\xit ``An N=1 superfield
action for M2 branes''}, \arxiv{0806}{2270}.}

\ref\CherkisSamann{S. Cherkis and C. S\"amann, {\xit ``Multiple
M2-branes and generalized 3-Lie algebras''}, \arxiv{0807}{0808}.}


\headtext={M. Cederwall: ``Superfield actions for N=8 and N=6...''}

\line{
\epsfxsize=18mm
\epsffile{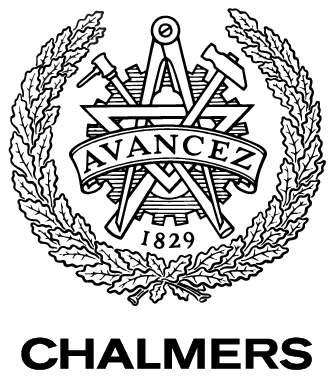}
\hfill}
\vskip-12mm
\line{\hfill G\"oteborg preprint}
\line{\hfill September, {\old2008}}
\line{\hrulefill}

\vfill
\vskip.5cm

\centerline{\sixteenhelvbold
Superfield actions for N=8 and N=6} 

\vskip4\parskip

\centerline{\sixteenhelvbold
conformal theories in three dimensions} 

\vfill

\centerline{\twelvehelvbold
Martin Cederwall}

\vfill

\centerline{\it Fundamental Physics}
\centerline{\it Chalmers University of Technology}
\centerline{\it SE 412 96 G\"oteborg, Sweden}

\vfill

{\narrower\noindent \underbar{Abstract:} The manifestly supersymmetric
pure spinor formulations
of the Bagger--Lambert--Gustavsson models with $N=8$ supersymmetry and the
Aharony--Berg\-man--Jafferis--Maldacena models with $N=6$
supersymmetry are given. The structures of the pure spinors are
investigated in both cases, 
and non-degenerate measures are formed using non-minimal
sets of variables, allowing for the formulation of an action principle. 
\smallskip}
\vfill

\font\xxtt=cmtt6

\vtop{\baselineskip=.6\baselineskip\xxtt
\line{\hrulefill}
\catcode`\@=11
\line{email: martin.cederwall@chalmers.se\hfill}
\catcode`\@=\active
}

\eject

\noindent There has recently been much interest in conformal
three-dimensional theories. Following the discovery of the existence
of a maximally supersymmetric ($N=8$) interacting theory of scalar
multiplets coupled to Chern--Simons, the Bagger--Lambert--Gustavsson
(BLG) theory [\BaggerLambertI,\Gustavsson,\BaggerLambertII,\BaggerLambertIII], 
much effort has been spent on trying to generalise the
construction and to interpret it in terms of an AdS boundary model of
multiple M2-branes. The interesting, but restrictive, algebraic
structure of the model, containing a 3-algebra with antisymmetric
structure constants, turned out to have only one finite-dimensional
realisation [\Papadopoulos,\GauntlettGutowski], 
possible to interpret in term of two M2-branes [\LambertTong,\DMPvR] (see
however refs. [\HoMatsuo,\BandosTownsend] 
dealing with the infinite-dimensional solution
related to volume-preserving diffeomorphisms in three dimensions).

It then became an urgent question how the stringent requirements in
the BLG theory could be relaxed. There are different
possibilities. One may let the scalar product on the matter
representation be degenerate [\GranNilssonPetersson]. 
This works at the level of equations
of motion, but does not allow for an action principle. One may also go
one step further, and add further null directions to that degenerate
case, which leads to scalar products with indefinite signature 
[\GomisMilanesiRusso,\BRTV,\HoImamuraMatsuo] (and
consequently to matter kinetic terms with different signs). Or, finally,
one may reduce the number of supersymmetries, specifically to $N=6$,
as proposed by Aharony, Bergman, Jafferis and Maldacena (ABJM)
[\ABJM], or maybe even to lower $N$ [\GaiottoWitten,\HLLLPI].
The $N=6$ models were further studied in refs. 
[\BKKS,\NishiokaTakayanagi,\MinahanZarembo,\HLLLPII,\BaggerLambertIV,\SchnablTachikawa,\NilssonPalmkvist]
(among other papers).  
For recent developments in the theory of multiple membranes, we refer
to ref. [\StringTalks] and references given there. The literature on
the subject is huge, and we apologise for omissions of references
to relevant papers.

The superfield
formulation of the BLG model was given in our previous paper 
[\CederwallBLG] (see also 
ref. [\BandosBLG], where the on-shell superfields were constructed for the
example of the BLG model based on the
infinite-dimensional algebra of volume-preserving diffeomorphisms in
three dimensions). A superfield formulation with $N=1$ superfields was given in
ref. [\MauriPetkou] and with $N=2$ superfields in ref. 
[\CherkisSamann]. In ref. [\CederwallBLG]  
we constructed an action in an $N=8$ pure spinor superspace
formulation of the BLG model, which covers all situations with $N=8$
above except the ones with degenerate scalar product. The purpose of
the present paper is twofold. Firstly, we construct the corresponding
formulation for $N=6$ superfields, thus covering the ABJM models. These
$N=6$ models are of course not maximally supersymmetric, but still more than
half-maximally, so the component actions have only on-shell
supersymmetry, which means that appropriate pure spinors are needed.
Secondly, in ref. [\CederwallBLG], some aspects of
the measure on non-minimal pure spinor space were left out, and simply
assumed to work in a similar way as in $D=10$. Here, we remedy this
omission by analysing the pure spinor constraints, adding
non-minimal variables in the spirit of ref. [\BerkovitsNonMinimal] and
forming explicit non-degenerate measures, both for $N=8$ and $N=6$,
thus completing the construction of manifestly supersymmetric actions
for the BLG and ABJM models, which can hopefully be used to improve on
quantum calculations [\GustavssonII,\BedfordBerman]. 

Let us first briefly review the results of ref. [\CederwallBLG]. Since the BLG
model is maximally supersymmetric, component formulations and also
usual superspace formulations are on-shell. There is no finite set of
auxiliary fields. A pure spinor treatment is necessary in order to
write an action in a generalised BRST setting. (For the use of pure
spinors and pure spinor superspaces in string theory we refer to
refs. [\BerkovitsI,\BerkovitsICTP,\BerkovitsNonMinimal], and in field
theory to
refs. [\NilssonPure,\HowePureI,\HowePureII,\CederwallNilssonTsimpisI,\CederwallNilssonTsimpisII,\BerkovitsParticle,\SpinorialCohomology,\Movshev,\CederwallNilssonSix,\CGNN,\CGNT,\HoweTsimpis].)  
The Lorentz algebra in $D=3$ is $so(1,2)\approx sl(2,\RR)$. The $N=8$
theory has an $so(8)$ R-symmetry, and we choose the fermionic
coordinates and derivatives to transform as $({\bf 2},{\bf
8}_s)=(1)(0010)$ under $sl(2)\oplus so(8)$. This representation is
real and self-conjugate. The pure spinors transform in the same
representation, and are written as $\l^{A\a}$, where $A$ is the
$sl(2)$ index and $\a$ the $so(8)$ spinor index. As usual, a BRST
operator is formed as $Q=\l^{A\a}D_{A\a}$, $D$ being the fermionic
covariant derivative. The nilpotency of $Q$ demands that
$$
(\l^A\l^B)=0\komma\Eqn\EightConstraint
$$
where $(\ldots)$ denotes contraction of $so(8)$ spinor indices, 
since the superspace torsion has to be projected
out. This turns out to be the full constraint\foot\star{The
vanishing of the ``torsion representation'' --- the vector part of the
spinor bilinear --- is necessary, but does not
always give the full pure spinor constraint. One example where further
constraints are needed is $N=4$, $D=4$ super-Yang--Mills theory.}. 
These pure spinors are similar to those encountered in
ref. [\FreGrassi]. The
``pure spinor wave function'' for the Chern--Simons field is a
fermionic scalar $\Psi$ of (mass) dimension 0 and ghost number 1. For the
matter multiplet we have a bosonic field $\Phi^I$ in the $so(8)$ vector
representation $(0)(1000)$ of dimension 1/2 and ghost number 0. In
addition to the pure spinor constraint, the matter field is identified
modulo transformations
$$
\Phi^I\rightarrow\Phi^I+(\l^A\sigma^I\varrho_A)\Eqn\PhiReducibility
$$ 
for arbitrary
$\varrho$. In
this minimal pure spinor formulation the fields are expanded in power
series in $\lambda$, \ie, in decreasing ghost number. The
field content (ghosts, fields and their antifields) are read off from
the zero-mode BRST cohomology given in tables 1 and 2 for the
Chern--Simons and matter sectors respectively.
\vskip2\parskip
\vbox{
$$
\vtop{\baselineskip20pt\lineskip0pt
\ialign{
$\hfill#\quad$&$\,\hfill#\hfill\,$&$\,\hfill#\hfill\,$&$\,\hfill#\hfill\,$
&$\,\hfill#\hfill\,$&$\,\hfill#\hfill\,$\cr
\hfill\hbox{gh\#}=&1    &0    &-1    &-2  &-3 \cr
\hbox{dim}=0&(0)(0000)&\phantom{(0)(0000)}&       &          \cr
        \Fr12&\bullet&\bullet&\phantom{(0)(0000)}   \cr
           1&\bullet&(2)(0000)&\bullet&\phantom{(0)(0000)} \cr
       \Fr32&\bullet&\bullet&\bullet&\bullet&\phantom{(0)(0000)}     \cr
           2&\bullet&\bullet&(2)(0000)&\bullet&\bullet\cr
       \Fr52&\bullet&\bullet&\bullet&\bullet&\bullet\cr
           3&\bullet&\bullet&\bullet&(0)(0000)&\bullet\cr
       \Fr72&\bullet&\bullet&\bullet&\bullet&\bullet\cr
}}
$$
\vskip2\parskip
\centerline{\it Table 1. The cohomology of the scalar complex.}
}

\vskip2\parskip
\vbox{
$$
\vtop{\baselineskip20pt\lineskip0pt
\ialign{
$\hfill#\quad$&$\,\hfill#\hfill\,$&$\,\hfill#\hfill\,$&$\,\hfill#\hfill\,$
&$\,\hfill#\hfill\,$&$\,\hfill#\hfill\,$\cr
\hfill\hbox{gh\#}=&0    &-1    &-2    &-3  &-4 \cr
\hbox{dim}=\Fr12&(0)(1000)&\phantom{(0)(0000)}&       &          \cr
        1&(1)(0001)&\bullet&\phantom{(0)(0000)}   \cr
           \Fr32&\bullet&\bullet&\bullet&\phantom{(0)(0000)} \cr
       2&\bullet&(1)(0001)&\bullet&\bullet&\phantom{(0)(0000)}     \cr
           \Fr52&\bullet&(0)(1000)&\bullet&\bullet&\bullet\cr
       3&\bullet&\bullet&\bullet&\bullet&\bullet\cr
           \Fr72&\bullet&\bullet&\bullet&\bullet&\bullet\cr
}}
$$
\vskip2\parskip
\centerline{\it Table 2. The cohomology of the vector complex.}
}
\vskip2\parskip

\noindent We observe that the field content is the right one. In
$\Psi$ we find the ghost, the gauge connection, its antifield and the
antighost. The antifield has dimension 2 (as opposed to \eg\ $D=10$
super-Yang--Mills, where it has dimension 3), indicating equations of
motion that are first order in derivatives. In $\Phi$ we find the
eight scalars $\phi^I$, the fermions $\chi^{A\dot\a}$ and their
antifields. In addition, the field $\Psi$ transforms in the adjoint
representation 
{\bf adj} of some gauge group and $\Phi^I$ in some representation {\bf
R} of the gauge group. The corresponding indices are suppressed. 

In ref. [\CederwallBLG], it was assumed that a non-degenerate measure
can be formed using a non-minimal extension of the pure spinor
variables along the lines of ref. [\BerkovitsNonMinimal]. This measure,
including the three-dimensional integration, should carry dimension 0
and ghost number $-3$, and should allow ``partial integration'' of the
BRST charge $Q$. It was then shown that the Lagrangian of the
interacting model is of a very simple form, containing essentially a
Chern--Simons like term for the Chern--Simons field, minimally coupled
to the matter sector:
$$
{{\cal L}}=\<\Psi,Q\Psi+\fr3[\Psi,\Psi]\>_{\xadj}
    +\fr2M_{IJ}\<\Phi^I,Q\Phi^J+\Psi\cdot\Phi^J\>_{\xR}\punkt\Eqn\Lagrangian
$$
The brackets denote (non-degenerate) scalar products on {\bf adj} and
{\bf R}, $[\cdot,\cdot]$ the Lie bracket of the gauge algebra and
$T\cdot x$ the action of the Lie algebra element in the representation
{\bf R}. $M_{IJ}$ is the pure spinor bilinear
$\e_{AB}(\l^A\sigma_{IJ}\l^B)$, which is needed for several reasons: in order to
contract the indices on the $\Phi$'s antisymmetrically, to get a
Lagrangian of ghost number 3, and to ensure invariance in the
equivalence classes defined by eq. (\PhiReducibility).

The invariances of the interacting theory, generalising the BRST
invariance in the linearised case, are:
$$
\eqalign{
\d\Psi&=Q\Psi-[\Lambda,\Psi]-M_{IJ}\{\Phi^I,\Xi^J\}\komma\hfill\cr
\d\Phi^I&=-\Lambda\cdot\Phi^I+(Q+\Psi\cdot)\Xi^I\komma\hfill\cr}
\eqn
$$
where $\Lambda$ is an adjoint boson of dimension 0 and ghost number 0,
and $\Xi^I$ a fermionic vector in {\bf R} of dimension 1/2 and ghost number
$-1$. Here we also introduced the bracket $\{\cdot,\cdot\}$ for the
formation of an adjoint from the antisymmetric product of two elements
in {\bf R}, defined via 
$\<x,T\cdot y\>_{\xR}=\<T,\{x,y\}\>_{\xadj}$. 
The invariance with parameter $\Lambda$ is manifest. The
transformation with $\Xi$ has to be checked. One then finds that the
transformation of the matter field $\Phi$ gives a ``field strength''
contribution from the anticommutator of the two factors $Q+\Psi$,
which is cancelled against the variation of the Chern--Simons
term. The single remaining term comes from the transformation of the
$\Psi$ in the covariant matter kinetic term, and it is proportional to 
$M_{IJ}M_{KL}\<\{\Phi^I,\Phi^J\},\{\Phi^K,\Xi^L\}\>_{\xadj}$. Due to the
pure spinor constraint, $M_{[IJ}M_{KL]}=0$, so if the structure
constants of the 3-algebra defined by
$\<\{x,a\},\{b,c\}\>_{\xadj}=\<x,\leftbr a,b,c\rightbr\>_{\xR}$ are
antisymmetric, this term vanishes. It was also checked that the
commutator of two $\Xi$-transformations gives a
$\Lambda$-transformation together with a transformation of the type
(\PhiReducibility).

Having thus reviewed the results of ref. [\CederwallBLG], we would
like to do the corresponding construction for $N=6$. 
The R-symmetry now is $so(6)\approx su(4)$.
We use $A_1\oplus A_3$ notations for Dynkin labels. The twelve
supercharges are in the (quasi-real) representation $(1)(010)$. The
four complex 
scalar fields should come in $(0)(100)$ (and their conjugates in
$(0)(001)$).
The pure spinor\foot\star{The representation of the fermionic
derivatives and of the $\l$'s are of course not spinor representations
of the R-symmetry group, only of the Lorentz group. For convenience,
we stick to the
terminology ``spinor'' and ``pure spinor'' also in this case.}
is $\l^{A\a\b}=-\l^{A\b\a}$, where $A=1,2$,
$\a,\b=1,\ldots4$. Later we will equivalently write $\l$ with an
$so(6)$ vector index as $\l^{Ai}$
The general symmetric product
of two ``spinors'' is
$\oplus_s^2(1)(010)=(0)(101)\oplus(2)(000)\oplus(2)(020)$. 
The second of these represents the torsion. 
We will need to keep the first one for writing
the matter lagrangian.
The pure spinor constraint is simply
$\e_{\a\b\g\d}\l^{A\a\b}\l^{B\g\d}=0$, or equivalently 
$$
\l^{Ai}\l^{Bi}=0\punkt\eqn
$$
It has the same formal structure as in the $N=8$ case, only that $\l$
is an $so(6)$ vector instead of an $so(8)$ vector (after triality rotation).

A scalar wave function has ``the same'' cohomology as in
ref. [\CederwallBLG] (in Table 1, just replace $(n)(0000)$ under $sl(2)\oplus
so(8)$ with $(n)(000)$ under $sl(2)\oplus su(4)$). So 
Chern--Simons is described in a formally identical manner. The matter
multiplet comes as expected from a bosonic wave function $\Phi^\a$ in
$(0)(100)$ and in the equivalence class
$$
\Phi^\a\approx\Phi^\a+\l^{A\a\b}\varrho_{A\b}\punkt\eqn
$$
The cohomology is the right one, shown in Table 3.

\vskip2\parskip
\vbox{
$$
\vtop{\baselineskip20pt\lineskip0pt
\ialign{
$\hfill#\quad$&$\,\hfill#\hfill\,$&$\,\hfill#\hfill\,$&$\,\hfill#\hfill\,$
&$\,\hfill#\hfill\,$&$\,\hfill#\hfill\,$\cr
\hfill\hbox{gh\#}=&0    &-1    &-2    &-3  &-4 \cr
\hbox{dim}=\Fr12&(0)(100)&\phantom{(0)(000)}&       &          \cr
        1&(1)(001)&\bullet&\phantom{(0)(000)}   \cr
           \Fr32&\bullet&\bullet&\bullet&\phantom{(0)(000)} \cr
       2&\bullet&(1)(001)&\bullet&\bullet&\phantom{(0)(000)}     \cr
           \Fr52&\bullet&(0)(100)&\bullet&\bullet&\bullet\cr
       3&\bullet&\bullet&\bullet&\bullet&\bullet\cr
           \Fr72&\bullet&\bullet&\bullet&\bullet&\bullet\cr
}}
$$
\vskip2\parskip
\centerline{\it Table 3. The cohomology of the $N=6$ matter complex.}
}
\vskip2\parskip

The field $\Phi^\a$ transforms in some representation ${\bf R}$ of the
gauge group, and $\bar\Phi_\a$ in $\bar{\bf R}$. 
The matter Lagrangian must again contain two powers of $\l$ through
the combination
$M_\a{}^\b=\fr2\e_{AB}\e_{\a\g\d\e}\l^{A\b\g}\l^{B\d\e}$, which is
exactly the $(0)(101)$. 
We write the Lagrangian:
$$
{{\cal L}}=\<\Psi,Q\Psi+\fr3[\Psi,\Psi]\>_{\xadj}
    +M_\a{}^\b\<\Phi^\a,(Q+\Psi\cdot)\bar\Phi_\b\>_{{\xR}\otimes\bar{\xR}}
    \Eqn\SixLagrangian
$$
with obvious notation. The generalised BRST invariance now reads
$$
\eqalign{
\d\Psi&=Q\Psi-[\Lambda,\Psi]-M_\a{}^\b\{\Phi^\a,\bar\Xi_\b\}
        -M_\a{}^\b\{\Xi^\a,\bar\Phi_\b\}\komma\hfill\cr
\d\Phi^\a&=-\Lambda\cdot\Phi^\a+(Q+\Psi\cdot)\Xi^\a\komma\hfill\cr}
\eqn
$$
The ``critical term'', as in the $N=8$ case, is the one that transforms the
$\Psi$ in the matter Lagrangian under the matter gauge
transformation. One gets a term proportional to
$$
M_\a{}^\g M_\b{}^\d\<\{\Phi^\a,\bar\Phi_\g\},
\{\Phi^\b,\bar\Xi_\d\}+\{\Xi^\b,\bar\Phi_\d\}\>_{\hbox{\sixbf adj}}\punkt\eqn
$$
Now, the tensor $N_{\a\b}{}^{\g\d}=M_\a{}^\g M_\b{}^\d$ 
turns out to be traceless and
symmetric in $(\a\b)$ and in $(\g\d)$, \ie, it transforms in
the 84-dimensional representation $(0)(202)$. 
This is the only $so(1,2)$ scalar at $\l^4$ due to the
pure spinor constraint. This gives a weaker condition on the
structure constants of the ``3-algebra'' than in the $N=8$ case: 
antisymmetry in pairs [\BaggerLambertIV], apart from the structure
already assumed. The classification of such algebraic structures was
performed in ref. [\SchnablTachikawa].
It is satisfactory that the structure of the pure spinors in both
cases give the necessary and sufficient algebraic structure by the
vanishing of a single term in the transformations. 

In ref. [\CederwallBLG], only the minimal pure spinors were
considered, and in practice regarded only as a book-keeping device
through the expansion in powers of $\l$. The existence of a
non-minimal extension of the variables along with a non-degenerate
measure was assumed in order that the action should be well-defined. 
We will now analyse the pure spinor constraints for the $N=8$ and
$N=6$ pure spinors, add non-minimal variables and show how the
non-degenerate measures of correct dimension and ghost number arise.

The measure is associated with the singlet cohomology of the
antighost in the Chern--Simons complex. With minimal pure spinor
variables, one may prescribe that this component of an integrand is
picked out, like a residue. Picking out a component at $\l^3\theta^3$
gives dimension 3 and ghost number $-3$, and together with the
three-dimensional $x$-integration dimension 0 and ghost number
$-3$. This goes well 
together with the Lagrangians above having dimension 0 and ghost
number 3. Such a ``measure'' is however degenerate, and can not be
used to form the actions, due to the fact that the fields are expanded
in positive powers of $\l$ only. 

A remedy, based on the analogous
construction in $D=10$ [\BerkovitsNonMinimal], is to introduce further
variables. Not only does the new measure become non-degenerate, it is
also defined in terms of full integrals over all variables, including
the $\theta$'s. Let us recall the 10-dimensional construction. In
addition to the pure spinor $\l^\a$ with the constraint
$(\l\g^a\l)=0$, one has another bosonic pure spinor $\m_\a$, with 
$(\m\g^a\m)=0$, of opposite chirality, 
and a fermionic spinor $r_\a$ fulfilling $(\m\g^ar)=0$.
We denote the
canonically conjugate variables (derivatives) to 
$\m_\a$ and $r_\a$ by $u^\a$ and
$s^\a$, respectively.  
The new BRST operator is ${\cal Q}=\l^\a D_\a+u^\a r_\a$, and its
cohomology is independent of $\m$ and $r$.
Let $\m$ have dimension $\hbox{dim}(\m)$ and ghost number 
$\hbox{gh\#}(\m)$. Then $r$ has dimension
$\hbox{dim}(\m)$ and ghost number $1+\hbox{gh\#}(\m)$.
In Euclidean signature, the pure co-spinor
$\m_\a$ can be seen as the complex conjugate of $\l^\a$.

In $D=10$, the pure spinor constraint is reducible, and has 5 independent
components, so a pure spinor has 11 (complex) degrees of freedom. The
same thing applies for the constraint on $r_\a$. 
The antighost singlet cohomology for $D=10$ super-Yang--Mills sits
at $\l^3\theta^5$, and is associated with a Lorentz invariant tensor
$T_{(\a_1\a_2\a_3)[\b_1\ldots\b_5]}$. There is of course a
corresponding tensor $\bar T^{(\a_1\a_2\a_3)[\b_1\ldots\b_5]}$ 
with conjugate indices. In
ref. [\BerkovitsNonMinimal] this tensor is used to form an invariant
integration measure for the pure spinor $\l$: 
$$
[d\l]\l^{\a_1}\l^{\a_2}\l^{\a_3}\sim{\star}\bar
T^{\a_1\a_2\a_3}{}_{\b_1\ldots\b_{11}}
d\l^{\b_1}\wedge\ldots\wedge d\l^{\b_{11}}
\komma\eqn
$$
where $\star$ refers to dualisation in the $\b$ indices.
We note that a requirement for this to work is that the number of
antisymmetric indices (five) equals the number of irreducible
constraints on the spinor, so that the integral is over the full pure
spinor space. The corresponding expression with conjugate indices
holds for the $\m$ integration, and for the $r$ integration we have
$$
[dr]\sim{\star}\bar T^{\a_1\a_2\a_3}{}_{\b_1\ldots\b_{11}}
\m_{\a_1}\m_{\a_2}\m_{\a_3}{\*\over\*r_{\b_1}}\ldots{\*\over\*r_{\b_{11}}}
\punkt\eqn
$$

Using these integration measures, and the ordinary ones for $x$ and
$\theta$, we list the dimensions and ghost numbers for the theory
after dimensional reduction to $D$ dimensions in Table 4.
So, the ghost numbers match, and also the dimensions (${1\over g^2}$
has dimension $D-4$ in $D$ dimensions), irrespectively of the
assignments of $\hbox{dim}(\m)$ and $\hbox{gh\#}(\mu)$.
 
\vskip3\parskip
\thicksize=\thinsize
\ruledtable
                |gh\#           |dim            \cr
$d^Dx$          |$0$            |$-D$             \crnorule
$d^{16}\theta$  |$0$            |$8$              \crnorule
$[d\l]$         |$8$            |$-4$             \crnorule
$[d\m]$         |$8\,\hbox{gh\#}(\mu)$   |$8\,\hbox{dim}(\m)$    \crnorule
$[dr]$          |$-11-8\,\hbox{gh\#}(\mu)$  |$-8\,\hbox{dim}(\m)$      \cr
total           |$-3$           |$-(D-4)$         
\endruledtable
\vskip2\parskip
\centerline{\it Table 4. The dimensions and ghost numbers of the
$D=10$ measure.}
\vskip2\parskip

The $\l$ and $\m$ integrations are non-compact and need
regularisation. In ref. [\BerkovitsNonMinimal] this is achieved,
following ref. [\MarneliusOgren], by the insertion of a factor 
$N=e^{\{{\cal Q},\chi\}}$. Since this differs from 1 by a ${\cal
Q}$-exact term, the regularisation is independent of the choice of the
fermion $\chi$. The choice $\chi=-\m_\a\theta^\a$ gives 
$N=e^{-\l^\a\m_\a-r_\a\theta^\a}$ and regularises the bosonic
integrations at infinity. At the same time, it explains how the term
at $\theta^5$ is picked out, this follows after integration over $r$.
$N$ has definite ghost number 0 if $\hbox{gh\#}(\mu)=-1$ and 
a dimensionful constant can be avoided in the regulator if 
$\hbox{dim}(\m)=\fr2$, so that $\hbox{gh\#}(r)=0$ and 
$\hbox{dim}(r)=\fr2$. 

In both the $N=8$ and $N=6$ theories in $D=3$, the na\"\i ve measure
sits at $\l^3\theta^3$. In analogy with the ten-dimensional case, we
need the number of irreducible constraints on the pure spinors to
equal the number of $\theta$'s. Indeed, the constraints, which in both
cases sit in the vector representation of $so(1,2)$, turn out to be
irreducible, which is straightforward to check. 
The pure spinor spaces are 13- and 9-dimensional,
respectively. In both cases, the spinor representation is the tensor
product of an $sl(2)$
doublet and a vector under an orthogonal group (in the $N=8$ case by
triality rotation). Letting $\l^1=a+ib$, $\l^2=c+id$, and choosing a
set of four orthogonal basis vectors $\{e_1,e_2,e_3,e_4\}$, the
general solution to the pure spinor constraint can be parametrised as
$$
\eqalign{
a&=\ell e_1\komma\cr
b&=\ell e_2\komma\cr
c&=\ell'(\sin\a\cos\b e_1+\sin\a\sin\b e_2+\cos\a e_3)\komma\cr
d&=\ell'(-\sin\a\sin\b e_1+\sin\a\cos\b e_2+\cos\a e_4)\punkt\cr
}\eqn
$$ 
There are four real parameters, and the stability group of the
parametrisation is $SO(N-4)\subset SO(N)$, so the real dimension of
pure spinor space is
$$
4+\hbox{dim}(SO(N))-\hbox{dim}(SO(N-4))=2(2N-3)\komma\eqn
$$ 
again giving (complex) dimensions 13 and 9 for the $N=8$ and $N=6$
cases, respectively. 

We can write the invariant tensors as 
$$
\eqalign{
&\e_{abc}(\l\g^a\theta)(\l\g^b\theta)(\l\g^c\theta)\cr
&\qquad=T_{(A_1\a_1,A_2\a_2,A_3\a_3)[B_1\b_1,B_2\b_2,B_3\b_3]}
\l^{A_1\a_1}\l^{A_2\a_2}\l^{A_3\a_3}
        \theta^{B_1\b_1}\theta^{B_2\b_3}\theta^{B_3\b_3}\cr}
\eqn
$$ in the $N=8$
        case, and as
$$
\eqalign{
&\e_{abc}(\l\g^a\theta)(\l\g^b\theta)(\l\g^c\theta)\cr
&\qquad=T_{(A_1i_1,A_2i_2,A_3i_3)[B_1j_1,B_2j_2,B_3j_3]}
\l^{A_1i_1}\l^{A_2i_2}\l^{A_3i_3}
        \theta^{B_1j_1}\theta^{B_2j_3}\theta^{B_3j_3}}
\eqn
$$ 
in the $N=6$ case (where in both cases the spinor contractions include
the $sl(2)$ index, and $\g^a$ are 3-dimensional $\g$-matrices).
The integration measure for a single $N=8$ pure spinor is then  
$$
[d\l]\l^{A_1\a_1}\l^{A_2\a_2}\l^{A_3\a_3}
\sim{\star}T^{A_1\a_1,A_2\a_2,A_3\a_3}{}_{B_1\b_1,\ldots,B_{13}\b_{13}}
d\l^{B_1\b_1}\wedge\ldots\wedge d\l^{B_{13}\b_{13}}\komma\eqn
$$
and for an $N=6$ pure spinor
$$
[d\l]\l^{A_1i_1}\l^{A_2i_2}\l^{A_3i_3}
\sim{\star}T^{A_1i_1,A_2i_2,A_3i_3}{}_{B_1j_1,\ldots,B_9j_9}
d\l^{B_1j_1}\wedge\ldots\wedge d\l^{B_9j_9}\punkt\eqn
$$
The same expressions apply for the $\m$ integrations, since the
``spinor'' representations in both cases are self-conjugate.
For the $r$ integrations we have
$$
[dr]\sim
{\star}T^{A_1\a_1,A_2\a_2,A_3\a_3}{}_{B_1\b_1,\ldots,B_{13}\b_{13}}
\m_{A_1\a_1}\m_{A_2\a_2}\m_{A_3\a_3}
{\*\over\*r_{B_1\b_1}}\ldots{\*\over\*r_{B_{13}\b_{13}}}
\eqn
$$
and
$$
[dr]\sim
{\star}T^{A_1i_1,A_2i_2,A_3i_3}{}_{B_1j_1,\ldots,B_9j_9}
\m_{A_1i_1}\m_{A_2i_2}\m_{A_3i_3}
{\*\over\*r_{B_1j_1}}\ldots{\*\over\*r_{B_9j_9}}
\eqn
$$
respectively.
Let us examine the dimensions and ghost numbers of the total
measures. The analogies of Table 4, with $\hbox{gh\#}(\mu)=-1$ and 
$\hbox{dim}(\m)=\fr2$, become
\vskip3\parskip
\ruledtable
\dbl\multispan2 \hfil$N=8$ \hfil     \dbl\multispan2 \hfil$N=6$\hfil\cr
               \dbl gh\#   |dim  \dbl gh\#    |dim    \cr
$d^3x$         \dbl$0$            |$-3$             
               \dbl$0$             |$-3$               \crnorule
$[d\theta]$    \dbl$0$            |$8$              
               \dbl$0$             |$6$               \crnorule
$[d\l]$         \dbl$10$            |$-5$            
               \dbl$6$             |$-3$               \crnorule
$[d\m]$         \dbl$-10$          |$5$    
               \dbl$-6$             |$3$               \crnorule
$[dr]$          \dbl$-3$            |$-5$      
               \dbl$-3$             |$-3$               \cr
total           \dbl$-3$           |$0$  
                \dbl$-3$           |$0$         
\endruledtable
\vskip2\parskip
\centerline{\it Table 5. The dimensions and ghost numbers of the
$N=8$ and $N=6$ measures.}
\vskip2\parskip
\noindent In both cases 
we get a non-degenerate measure of dimension 0 and ghost
number $-3$, as desired for a conformal theory. 
Also here, the measures of course have to be
regularised in the same way as in ref. [\BerkovitsNonMinimal]. We
insert a factor $N=e^{\{{\cal Q},\chi\}}$, where 
$\chi=-\m_{A\a}\theta^{A\a}$ for $N=8$, and
$\chi=-\m_{i\a}\theta^{i\a}$ for $N=6$. 

To conclude, we have extended our previous manifestly supersymmetric
formulation of the $N=8$ BLG models to the $N=6$ ABJM models. We have
also performed a detailed analysis of the pure spinor constraints and
provided proper actions based on non-degenerate measures on
non-minimal pure spinor spaces. We hope that these formulations may be
helpful in the future, \eg\ for the investigation of quantum
properties of the models.

\acknowledgements The author would like to thank Bengt E.W. Nilsson,
Ulf Gran, Dimitrios Tsimpis, Nathan Berkovits and
Pietro Antonio Grassi for discussions and comments.

\refout

\end